\documentclass[journal]{IEEEtran}
\usepackage{epsfig,amsfonts,amsbsy,bm,mathrsfs}
\usepackage{mathrsfs} 
\usepackage{graphicx,cite,amssymb,amsmath,bm}
\usepackage{mathtools}
\usepackage{amsmath} 
\usepackage{etex}
\usepackage{amsmath,amssymb}
\usepackage{lipsum}
\usepackage{array,verbatim}
\usepackage{makecell}
\usepackage{siunitx}
\usepackage{float}
\usepackage{xcolor}
\usepackage{colortbl}
\usepackage{tikz}
\usepackage{ctable}
\usepackage[english]{babel}
\usepackage{color}
\usepackage[utf8]{inputenc}
\usepackage[T1]{fontenc}
\usepackage{balance}
\usepackage{balance}
\usepackage{perpage}
\usepackage{algorithm}
\usepackage[noend]{algpseudocode}
\usepackage{subcaption}
\usepackage{graphbox,graphicx}
\usepackage{xcolor}
\usepackage{bbm}

\addto\captionsenglish{}
\makeatletter

\newcommand{\SNR}{\text{SNR}}

\newcommand{\xii}{\boldsymbol{x}^\text{I}}
\newcommand{\xq}{\boldsymbol{x}^\text{Q}}

\renewcommand{\a}{\boldsymbol{a}}
\renewcommand{\b}{\boldsymbol{b}}
\renewcommand{\u}{\boldsymbol{u}}
\newcommand{\s}{\boldsymbol{s}}

\newcommand{\ua}{\boldsymbol{u}^{{k}}}
\newcommand{\us}{\boldsymbol{u}^{\gamma n}}

\newcommand{\ent}{\mathsf{H}}
\newcommand{\Pe}{P_\mathsf{e}}
\newcommand{\rateHDD}{\mathsf{R}_{\mathsf{HDD}}}

\newcommand{\Rs}{R}
\newcommand{\nc}{\tilde{n}}

\newcommand{\kc}{\tilde{k}}

\newcommand{\cc}{\boldsymbol{C}}

\definecolor{OliveGreen}{cmyk}{0.64,0,0.95,0.40}

\newcommand{\SH}{\textcolor{black}}

\ifCLASSINFOpdf
\else
\fi

\begin{document}
%
\title{On Product Codes with Probabilistic Amplitude Shaping for High-Throughput Fiber-Optic Systems}

%
%
%

\author{Alireza~Sheikh \IEEEmembership{Member, IEEE}, Alexandre~Graell~i~Amat, \IEEEmembership{Senior Member, IEEE}, \\ and Alex Alvarado, \IEEEmembership{Senior~Member, IEEE} 
        \thanks{The work of A. Sheikh and A. Alvarado has received funding from the European Research Council (ERC) under the European Union's Horizon 2020 research and innovation programme (grant agreement No 757791).}
        \thanks{A. Sheikh and A. Alvarado are with the Department of Electrical Engineering, Eindhoven University of Technology, 5612 AZ Eindhoven, The Netherlands (emails: \{a.sheikh,a.alvarado\}@tue.nl).}
            \thanks{A. Graell i Amat is with the Department of Electrical Engineering, Chalmers University of Technology, SE-41296 Gothenburg, Sweden (email: alexandre.graell@chalmers.se).}
}

\maketitle

\begin{abstract}

Probabilistic amplitude shaping (PAS) can flexibly vary the spectral efficiency (SE) of fiber-optic systems. In this paper, we demonstrate the application of PAS to bit-wise hard decision decoding (HDD) of product codes (PCs) by finding the necessary conditions to select the PC component codes.  We show that PAS with PCs and HDD yields gains up to $2.7$ dB and SE improvement up to approximately $1$ bit/channel use compared to using PCs with uniform signaling and HDD. Furthermore, we employ the recently introduced iterative bounded distance decoding with combined reliability of PCs to improve performance of PAS with PCs and HDD.

\end{abstract}


\thispagestyle{empty} \setcounter{page}{0}


\section{Introduction}\label{int}

\IEEEPARstart{T}{o} keep up with the current trend on data demand and realize high spectral efficiencies (SEs), coding in combination with a high order modulation, a scheme known as coded modulation (CM), has become indispensable in fiber-optic communications. For instance, a recently commercially available optical transponder employs up to $64$ quadrature amplitude modulation (QAM) to provide $400$ Gbit/s and $600$ Gbit/s throughput on a single frequency channel, i.e,. $400$ Gbit/s/$\lambda$ and $600$ Gbit/s/$\lambda$, respectively\cite{fujitsu2019}. 

In practice, there is always a gap between the performance of CM and the corresponding theoretical limit. This gap is mainly due to two reasons: (i) the employed codes have finite length and suboptimal decoders are usually used in order to constrain the receiver complexity\SH{;} (ii) equidistant signal constellation points (e.g., square QAM) with uniform signaling are usually employed, yielding the asymptotic $1.53$ dB gap to the Shannon limit for transmission over the additive white Gaussian noise (AWGN) channel.  

Product-like codes such as product codes (PCs) \cite{Elias1954} and staircase codes \SH{(SCCs)} \cite{staircase_frank} with hard decision decoding (HDD) are particularity suitable for very high-throughput applications such as next generation fiber-optic systems  \cite{ITUT}, \cite{400GZR}. To improve the performance of CM with product-like codes, one approach is to boost the decoding performance. As HDD yields a considerable performance loss with respect to soft decision decoding (SDD), recently, several hybrid decoding schemes have been proposed in order to partially close the gap between HDD and SDD with limited additional complexity to that of HDD \cite{optimal_decsheikh,She19Tcom,She19,FougstedtiBDDSR2018,Montorsi2019,Yibitflip,Ksch18,Liga2019bit}.
In particular, iterative bounded distance decoding with combined reliability (iBDD-CR) recently introduced in \cite{optimal_decsheikh} is a low-complex hybrid decoding scheme for PCs which utilizes the channel reliabilities to improve the performance of HDD. 

The performance of CM can also be improved by means of probabilistic shaping \cite{Calderbank_1990,Forney_1992,georg_tcom}.
Probabilistic amplitude shaping (PAS) was proposed in \cite{georg_tcom} and attracted a significant interest in the fiber-optic community \cite{Buchali_2016,Sheikh2017Shap,boherer20191,Cho2019}. \SH{The original PAS architecture employs binary low-density parity check (LDPC) codes. However,} PAS with PCs and HDD/hybrid decoding can provide a solution for high-throughput fiber-optic systems.

In this paper, we apply PAS
to bit-wise HDD of PCs. The contribution is twofold: (i) 
We define an interleaver in the PAS architecture \cite{georg_tcom} that distributes the different bit-levels equally among the
	PC component codewords. Due to the structure of PCs, this can be accomplished by a random interleaver. Further, the 
structure of PCs  imposes some constraints on the PCs to be used with PAS. We find the necessary conditions on the parameters of  the PC component codes  in order to be used in combination with PAS;
(ii) we employ iBDD-CR decoding of PCs \cite{optimal_decsheikh} to improve the performance of PAS with PCs and (conventional) iterative bounded distance decoding (iBDD). We show that PAS with PCs and iBDD provides gains up to $2.7$ dB and a spectral efficiency (SE) improvement of up to $1$ bit/channel use (bpcu) compared the baseline scheme with uniform signaling. \SH{Furthermore, we show that PAS with PCs and iBDD-CR decoding can close the gap between PAS with iBDD of SCCs and PCs.}\footnote{\textbf{Notation}: 
\SH{$P_X(\cdot)$ denotes the probability mass function (PMF) of the RV $X$. Boldface letters denote vectors and matrices. ${x}_{i}$ is the $i$-th element of $\boldsymbol{x}$. $\mathbb{E}_X(\cdot)$ and $\ent(X)$ stand for the expectation and entropy of the RV $X$.}}


\section{Preliminaries}\label{GMIdis}

\begin{figure*}[!t]
	\centering
	\vspace{-0.4cm}
	\scalebox{0.7}{
		\includegraphics[scale=1.01]{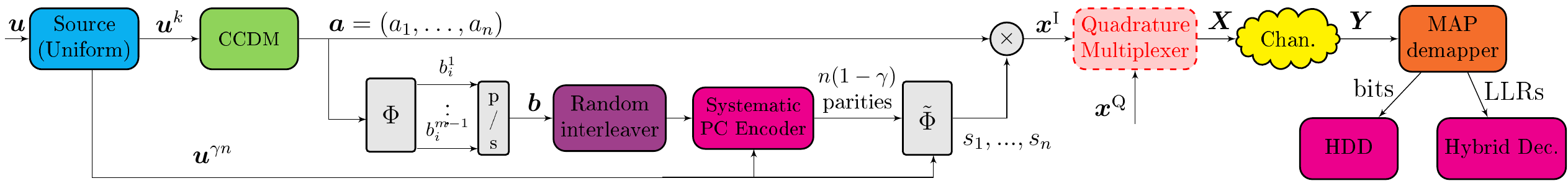}
	}
	\caption{Block diagram of the CM scheme with PAS and PC under consideration. The parameters shown in the system model corresponds to encoding \SH{and decoding} of one PC  code array. \SH{The HD demapper outputs bits and log-likelihood ratios (LLRs) corresponding to the PC code bits  for HDD and hybrid decoding, respectively.} 
	}
	\label{figsystemmodel}
	\vspace{-1.3ex}
\end{figure*}

\SH{Without loss of generality, we consider ASK modulation. However, in Sec.~\ref{sec:simulation} we report results for square QAM, which can be seen as the Cartesian product of two ASK constellations. The transmitted constellation points are chosen from $\mathcal{X}\triangleq\{-2^m+1,...,-1,1,...,2^m-1\}$, where $m$ is the number of bits per symbol.}
We consider an  AWGN channel.\footnote{The fiber-optic channel can be modeled as an AWGN channel using the Gaussian noise model \cite{Poggiolini2015}.} The channel output at time instant \emph{i} is given as
\begin{align}
\label{in_outAWGN}
{Y_i} = {X_i} + {Z_i} \qquad\qquad i=1,2,\ldots,n
\end{align}
where $n$ is the block length, $Z_i$ are independent and identically distributed (i.i.d.) Gaussian RVs with zero mean and unit variance, and ${X_i}$ is the \SH{constrained channel input, hence, $\SNR=\mathbb{E}_X\left[X^{2}\right]$}. We consider a block-wise transmission system where $\u$ stands for the transmitted information block and $\hat \u$ denotes the corresponding decoded block.  

Similar to \cite{Sheikh2017Shap}, in this paper we consider the PAS scheme of \cite{georg_tcom} with a binary code for transmission and bit-wise HDD, i.e., bit-wise Hamming metric decoding, at the receiver side. The achievable information rate of such a system, denoted by $\rateHDD$, is derived in \cite[eq.~(62)]{bocherer2017}. Similar to \cite{georg_tcom,Sheikh2017Shap}, we also consider the Maxwell-Boltzmann distribution with shaping parameter $\lambda$, where the PMF of a constellation point $x \in \mathcal{X}$ is given as ${P^{\lambda}_{X}}\left( x \right) = \frac{\exp\left( - \lambda x^2\right)}{\sum\limits_{\tilde{x} \in \mathcal{X}} \exp\left( - \lambda \tilde{x}^2\right) }$.
For each SNR, $\lambda$ is optimized such that $\rateHDD$ is maximized, i.e.,  
\begin{align}
\label{MI_rate}
\lambda^{*}=\mathop {{\text{argmax}}}\limits_{\lambda > 0} \;\rateHDD.
\end{align}
We use a maximum a-posteriori (MAP) \SH{demapper}, which employs $P_X^{{\lambda ^*}}$ for the detection.

\section{Coded Modulation with PAS and PCs}\label{coded_modulation}

The idea of PAS is to shape the constellation amplitudes and then utilize the parities of a systematic code for the sign bits. Let us denote by $ \mathcal{A}\triangleq \{1,3,\dots,2^m-1\}$ the set of ASK amplitudes, of size $2^{m-1}$.  For a target $P_X^{{\lambda ^*}}$, the PMF of an amplitude $a \in \mathcal{A}$ is $P^{\lambda^{*}}_{A}\left( a \right) = 2P_X^{{\lambda ^*}}\left( {a} \right)$, as the sign bis are uniformly distributed. 
The schematic of PAS considered in this paper is shown in Fig.~\ref{figsystemmodel}. In the following, we briefly review the different components of  Fig.~\ref{figsystemmodel}. In what follows, the PAS input is a vector of bits of length $k+\gamma n$, where $k$ bits are employed for amplitude shaping. Furthermore, the output of PAS is a vector of shaped symbols of length $n$, where each symbol is taken from an ASK constellation of size $2^m$.

Let $\u=(u_1,u_2,\dots,u_{k+\gamma n})$ be the information vector of length $k+\gamma n$ bits with uniformly distributed components, where $u_i \in \{0,1\}$ for $i  = 1,2,\ldots,k+\gamma n$. The vector $\u$ is parsed to $\us$ and $\ua$, of lengths $\gamma n$ and $k$, respectively. The vector $\ua$ is the input to the shaping block, which generates a sequence of amplitudes $\a=(a_1,a_2,...,a_n)$ with distribution $P^{\lambda^{*}}_{A}$. 
The shaping block employs distribution matching. 
The block length considered in this paper is large (in the order of $12$k--$260$k), hence, there is no difference between the performance of distribution matchers \cite{shapingGeorg,Fehenberger2018multi}. Thus, we use the constant composition distribution matching (CCDM) method proposed in \cite{shapingGeorg}. We also consider the binary reflected Gray coding (BRGC) at the mapper $\Phi$. In particular, $\Phi$ generates $m-1$ bits corresponding to binary image of shaped amplitudes, yielding sequence $\b=(\b_1,\b_2,\ldots,\b_n)$ of length $n(m-1)$ bits, where $\b_i=(b^{1}_{i},b^{2}_{i},\ldots,b^{m-1}_{i})$ for $i = 1,2,\ldots,m-1$. The sequence $\b$ and $\us$ are interleaved yielding a sequence of length $n(m-1)+n\gamma$, which is encoded  using  systematic PCs. 
The encoding generates parity bits of length $n(1-\gamma)$. 
By combining $n(1-\gamma)$ parity bits with $\us$ and employing the mapping according to $0\mapsto +1$ and $1\mapsto -1$ in block $\tilde \Phi$, the uniformly distributed sign sequence $\s=(s_1,s_2,\ldots,s_n)$ of length $n$ is yielded. The element-wise multiplication of $\a$ by $\s$ generates a sequence \SH{$\xii=(x^\text{I}_1,x^\text{I}_2,\ldots,x^\text{I}_n)$} with the desired distribution \SH{for the ASK symbols}\footnote{\SH{For QAM modulation, $\xq$ is generated similarly and multiplexed with $\xii$, which yields a sequence with the desired distribution for the QAM symbols.}}. 
The \SH{hard detected (HD) sequence} at the receiver is first decoded and then de-interleaved, resulting in the sequences $\hat{\a}=(\hat{a}_1,\hat{a}_2,\ldots,\hat{a}_n)$  and $\hat{\boldsymbol{u}}^{k}$. Finally $\hat{\a}$ is de-shaped using an inverse CCDM and combined with $\hat{\boldsymbol{u}}^{k}$ to generate $\hat{\boldsymbol{u}}$ as the decision on ${\boldsymbol{u}}$. In the following, we discuss the encoding, interleaving, and decoding in more details.

\begin{figure*}[t]
		\begin{minipage}[t]{0.3\linewidth}
		\centering
		\scalebox{0.65}{		
			\includegraphics[]{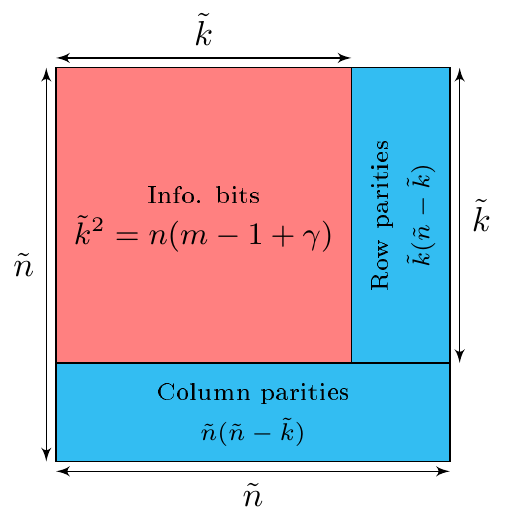}
		}
		\vspace{-1ex}
		\caption{Code array of a PC.}
		\label{staircase_code}
	\end{minipage}
	\begin{minipage}[c]{0.7\linewidth}
	\renewcommand{\tabcolsep}{0.15cm}
\vspace{-1.2cm}
	\begin{center}
			\captionof{table}{Parameters of the designed PCs for $v=10$, $t=3$, and $16$-ASK modulation}
			\vspace{-1ex}
					\scalebox{0.82}{
			\begin{tabular}{cccccccccccc}
			\arrayrulecolor{black}\hline
			\toprule
			$\gamma$ & $0.7682$ & $0.7503$ & $0.6912$ & $0.6797$  &  $0.5942$ &   $0.5233$  &  $0.4464$  &  $0.3645$  &  $0.2303$  &  $0.1426$ &  $0.00854$
			\\
			\midrule
			$s$ & $3$  &  $77$ &  $261$ &  $289$  & $447$ &  $535$ &  $605$  & $661$  & $727$ &  $759$ &  $797$
			\\
			\midrule				
			$\nc$ & $1020$    &     $946$    &     $762$    &     $734$   &      $576$     &    $488$   &      $418$   &      $362$      &   $296$    &     $264$     &    $226$ \\				
			\midrule				
			$\kc$ &  $990$ &  $916$  & $732$ &  $704$ &  $546$ &  $458$ &  $388$ &  $332$  & $266$ &  $234$ &  $196$  \\								
			\midrule				
			$n$ & $260100$  &    $223729$  &    $145161$  &    $134689$  &     $82944$    &   $59536$   &    $43681$   &    $32761$   &    $21904$   &    $17424$   &    $12769$	 \\
			\midrule				
			$\gamma n$ & $199800$  &  $167869$   &   $100341$    &   $91549$   &   $49284$   &   $31156$    &  $19501$   &    $11941$    &    $5044$    &    $2484$     &    $109$	
			\\			
			\midrule				
			$\Rs$ & $0.9420$ &   $0.9376$  &  $0.9228$  &  $0.9199$  &  $0.8985$   & $0.8808$ &   $0.8616$  &  $0.8411$  &  $0.8076$  &  $0.7856$   & $0.7521$
			\\		
			\hline
			\toprule                			
				\label{table:feasibleparam}	\end{tabular}}\end{center}
\vspace{1cm}	
	\end{minipage}
\vspace{-1.8cm}
\end{figure*}

We consider binary PCs with Bose-Chaudhuri-Hocquenghem (BCH) component codes. Let $\mathcal{C}$ be a systematic BCH component code of length $\nc$ and information length $\kc$, in short $(\nc,\kc)$. $\mathcal{C}$ is constructed over the Galois field $\text{GF}(2^v)$ with (even) block length $\nc$ and information block length $\kc$ given by
\begin{align}
& \nc=2^v-1-s, \label{n_formula}\\
& \kc=2^v-vt-1-s \label{k_formula}
\end{align}
where $s$ and $t$ are the shortening length and the error correcting capability of $\mathcal{C}$, respectively. Therefore, a shortened BCH code is completely specified by the parameters $(v,t,s)$. A PC with $(\nc,\kc)$ component codes is defined as the set of all $\nc \times \nc$ arrays $\cc=[c_{i,j}]$ such that each row and column of $\cc$ is a valid codeword of $\mathcal{C}$. Fig.~\ref{staircase_code} shows the code array of a PC. The red part corresponds to the information bits while the parity bits are shown in blue.

In a conventional CM with PCs, for  given component parameters $v$ and $t$, $s$ can be selected without any constraints. However, the PAS architecture imposes a constraint on the parameters of the component codes of the PC, i.e., for a given $v$ and $t$, only some values of $s$ are feasible. In what follows, we find such constraint. 
Let us consider a PC with $(\nc,\kc)$ BCH component code corresponding to a code rate  ${\Rs} = \frac{\kc^2}{\nc^2}$. 
Using a $2^m$-ASK constellation in the PAS, $\Rs$ is given as
\begin{align}
\label{componentcoderate}
\Rs = \frac{\kc^2}{\nc^2} =  \frac{{m - 1 + \gamma }}{m},
\end{align} 
where $0 \le \gamma  < 1$ is a tuning parameter used to vary the SE of PAS. We assume that the bit sequence corresponding to the output of the interleaver (of size $(m-1+\gamma)n$) is parsed into $\kc$ sequences of length $\kc$, each placed as information bits corresponding to the $i$-th row ($i = 1,\ldots,\kc$) of the PC code array (see red part in  Fig.~\ref{staircase_code}). Therefore,  $n$ should be selected such that
\begin{align}\label{parity}
n=\frac{\kc^2}{m-1+\gamma}.
\end{align}
As explained before, the number of parity bits should be $n(1-\gamma)$. Employing \eqref{componentcoderate} and \eqref{parity}, the total number of parity bits corresponding to a PC codeword is given as 
\begin{align}
\label{eq:ncmkc}
\nc^2-\kc^2=\kc^2\left(\frac{1}{\Rs}-1\right)={n(1-\gamma)}.
\end{align}  
The component codes of the PC employed in the PAS should then be chosen such that the block length \eqref{parity} and the number of parity bits \eqref{eq:ncmkc} are positive and non-negative integers, respectively, i.e., 
\begin{align}
{n}\in \{1,2,\ldots\}, \; \text{and} \;
{\gamma n}\in\{0,1,\ldots\}.\label{eq:cond3}
\end{align} 
Therefore, in order to use PCs with component code parameters $(v, t)$ in PAS with  $2^{m}$-ASK constellation, the feasible values of $s$ must satisfy \eqref{parity} and \eqref{eq:cond3}. We highlight that a feasible value of $s$ corresponds to a feasible value of $\gamma$ (by employing \eqref{n_formula}--\eqref{k_formula} in \eqref{componentcoderate}, one can see a one-to-one relation between $s$ and $\gamma$). 

For $16$-ASK modulation ($m=4$), $v=10$ and $t=3$, a subset of all feasible values for $s$ and correspondingly $\gamma$, $(\nc,\kc)$, $n$, and $\Rs$ are given in Table~\ref{table:feasibleparam}. The same table for SCCs with $16$-ASK modulation ($m=4$), $v=10$, and $t=3$ was given in \cite[Table~II]{Sheikh2017Shap}. Comparing Table~\ref{table:feasibleparam} with \cite[Table~II]{Sheikh2017Shap} for a given code rate, one can see that the structural difference between PCs and SCCs yields different component code parameters for each code. We highlight that the feasible values for $s$ (and thus $\gamma$) in Table~\ref{table:feasibleparam} are only a subset of a total of $205$ feasible values. One can check that the total number of feasible values for $s$ (and thus $\gamma$) for a SCC is only $40$ (see \cite[Table~II]{Sheikh2017Shap}). As it is shown in Sec.~\ref{sec:simulation}, by varying $\gamma$ one can change the SE of PAS. Therefore, PAS with PCs provides a much finer granularity of SEs than PAS with SCCs, due to the larger number of feasible values of $\gamma$.

We assume that the sign bits generated by PAS, i.e., $(s_1,\ldots,s_n)$,  represent the first bit level in the BRGC label of the transmitted symbol. For instance, the transmitted signal points with the corresponding BRGC labels for 8-ASK modulation are shown in Table~\ref{labeling}. It is well-known that the reliability of different bit levels varies in CM. For symmetric PCs, i.e., with the same row and column component codes, the performance of iterative decoding  improves if both row and column component decoder observe the same channel. This can be achieved by uniformly spreading the bits of $\boldsymbol{b}$ and $\us$ (which correspond to different bit levels) between the component codes of the PC. With this aim, we employ a random interleaver in the PAS architecture (see Fig.~\ref{figsystemmodel}).
We highlight that there is no random interleaver in \cite{Sheikh2017Shap}. In fact, a simple structured interleaver is considered in \cite[Sec.~IV-C]{Sheikh2017Shap} for SCCs, which ensures that bits from both $\boldsymbol{b}$ and $\us$ are uniformly distributed between rows in each staircase block.

	\begin{table}[t]
		\caption {BRGC of the amplitudes for 8-ASK} \label{T2}
		\vspace{-0.09cm}
		\begin{center}
			\vspace{-2ex}
			\begin{center}\begin{tabular}{ccccccccc}
					\arrayrulecolor{black}\hline
					\toprule

					amplitude & $-7$ & $-5$ & $-3$ & $-1$ & $1$  & $3$ & $5$ & $7$ \\
					\midrule
					label  &  $110$ & $111$ & $101$ & $100$ & $000$  & $001$ & $011$ & $010$ \\
					\hline
					\toprule                			
			\end{tabular} \end{center}
		\end{center}
	    \label{labeling}
		\vspace{-0.6cm}
\end{table}
	
PCs are usually decoded using iBDD of the component codes \cite{Justesen2011}. In this paper, we also consider the iBDD-CR algorithm introduced in \cite{optimal_decsheikh} for PCs. In order to apply iBDD-CR in PAS with PCs, the LLR of the hard detected bits at the decoder should be computed. 
Given the received symbol ${Y_i}$ (see \eqref{in_outAWGN}), one can compute the LLR corresponding to the $l$-th  bit ($l =  1,\ldots,m $) of the binary image of $Y_{i}$ as 
\begin{align} \label{LLR}
L_i^l = \ln \left(\frac{\sum\limits_{\tilde{x}_i \in \mathcal{S}_l^0} {{e^{ - \frac{{({Y_i} - {\tilde{x}_i}{)^2}}}{{2}}}}}\cdot{P^{\lambda^{*}}_{X}}\left( \tilde{x}_i \right) }{{\sum\limits_{{\tilde{x}_i} \in \mathcal{S}_l^1} {{e^{ - \frac{{({Y_i} - {\tilde{x}_i}{)^2}}}{{2}}} \cdot{P^{\lambda^{*}}_{X}}\left( \tilde{x}_i \right)  }} }}\right),
\end{align}  
where $\mathcal{S}_l^0$ and $\mathcal{S}_l^1$ are sets of size $2^{m}$ ASK symbols with $0$ and $1$ as the $l$-th bit of the corresponding BRGC label, respectively. 

 \begin{figure}[!t] \centering 
\includegraphics[width=\columnwidth]{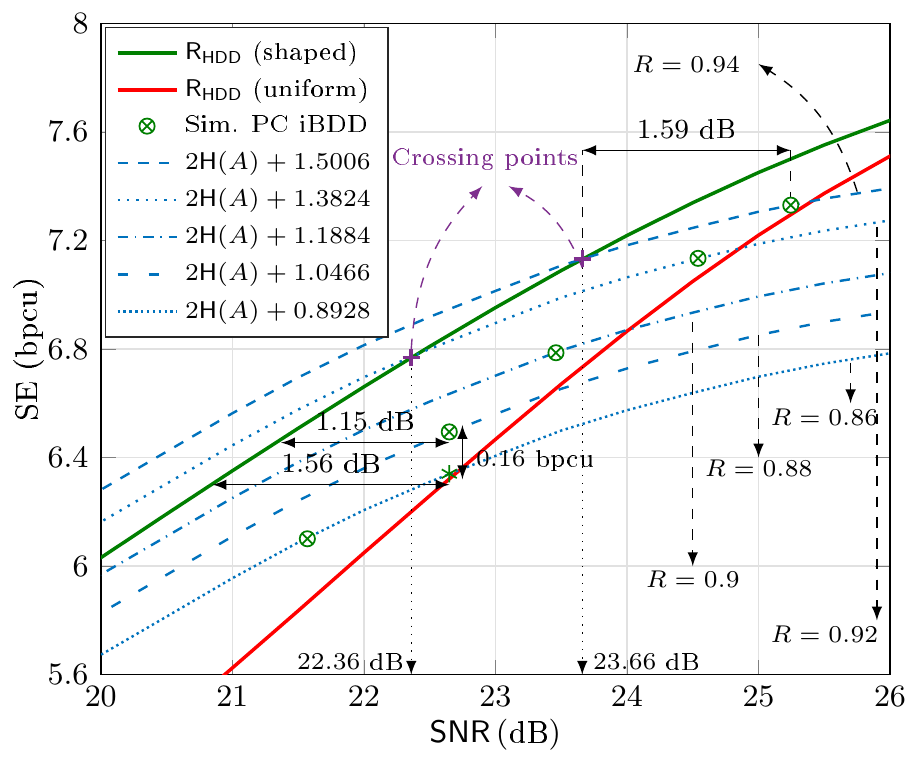}  
	\vspace{-4ex}
	\caption{Simulation results of the PAS based on PCs with $s=77$, $261$, $447$, $535$, $605$, for $256$-QAM, and comparison with the corresponding achievable information rates.} \vspace{-2ex}
	\label{16_ASK_sim} 
\end{figure}

\section{Numerical Results}\label{sec:simulation}

We evaluate the performance of PAS using PCs with parameters given in Table~\ref{table:feasibleparam} for transmission of \SH{$256$-QAM}. Note that the Cartesian product of two $16$-ASKs with distribution $P_X^{{\lambda ^*}}$ gives the shaped $256$-QAM.
The target block error probability ($\Pe$) is set to $10^{-3}$.
The decoding of the PC is performed using a maximum of $8$ decoding iterations employing  iBDD with extrinsic message passing \cite[Algorithm 1]{h_braided} and iBDD-CR \cite{optimal_decsheikh}. 

For the feasible values of $s$ given in Table~\ref{table:feasibleparam},
$n$ is large, hence the information rate of PAS for each quadrature is $\ent(A) + \gamma$. Therefore, the total rate is $2\ent(A) + 2\gamma$. In Fig.~\ref{16_ASK_sim}, we depict $\rateHDD$ (both shaped and uniform) for a $256$-QAM and $2\ent(A) + 2\gamma$ for values of $\gamma$ corresponding to $s=77$, $261$, $447$, $535$, and $605$. As can be seen, for $\gamma=0.4464$ ($s=605$) the curve $2\ent(A) + 0.8928$ is below the shaped $\rateHDD$, hence, employing a single PC with code  $R=0.8616$ (see Table~\ref{table:feasibleparam}, eighth column), one can achieve all points on $2\ent(A) + 0.8928$ by only changing the distribution of amplitudes. A similar observation can be made for $\gamma=0.5233$ ($s=535$) and $\gamma=0.5942$ ($s=447$). However, for $\gamma=0.7503$ ($s=77$) and $\gamma=0.6912$ ($s=261$) there are crossing points which are shown with plus signs. The crossing point between the curves $2\ent(A) + 2\gamma$ and shaped $\rateHDD$ specifies the optimal operating point, which can be achieved with infinity long block length codes and optimal decoding \cite{georg_tcom}. As suboptimal iterative decoding of PCs is considered in this paper, only the points on the curve $2\ent(A)+1.5006$ and $2\ent(A)+1.3824$ corresponding to SNRs larger than $23.66$ dB and $22.36$ dB, respectively, are achievable. Therefore, the SNRs larger than $23.66$ dB and $22.36$ dB defines the feasible SNR regions for PAS based on employing PCs with $s=77$ and $s=261$, respectively. 

We \SH{have simulated} the performance of the PAS with the considered PCs and then find the minimum SNR in the feasible SNR region where $\Pe$ can be achieved. In what follows, we refer to the point on the curve $2\ent(A) + 2\gamma$  corresponding to the minimum SNR achieving $\Pe$ as an \SH{\emph{operating point}}. In Fig.~\ref{16_ASK_sim}, we plot the operating points of the PAS corresponding to PCs with $s=77$, $261$, $447$, $535$, and $605$, for $256$-QAM to achieve $\Pe=10^{-3}$ (green circle with crosses). As it can be seen,  for $\gamma=0.7503$ (corresponding to crossing point with $\SNR=23.66$ dB), one should back-off roughly $1.59$ dB from the optimal operating point (plus sign corresponding to $\SNR=23.66$ dB) to achieve $\Pe=10^{-3}$. Note that the simulation points are just a subset of size $5$ out of $205$ feasible operating points  (see Sec.~\ref{coded_modulation}). 

We also notice that the slope of $2\ent(A) + 2\gamma$ is less than that of shaped $\rateHDD$, hence, we \SH{will} switch to another code rate (different $\gamma$) in order to operate as close as possible to shaped $\rateHDD$.
For instance, the point shown with asterisk is also on the achievable rate curve $2\ent(A)+0.8928$, corresponding to $\gamma=0.4464$. However, switching to $\gamma=0.5233$ (operating on the curve $2\ent(A)+1.0466$) reduces the gap between the operating point and shaped $\rateHDD$ by $0.41$ dB, which corresponds to $0.16$ bpcu higher SE for the same $\SNR$. 
Finally, Fig.~\ref{16_ASK_sim} shows that the achieved rates by the PAS with PCs are larger than the achievable rate of the (conventional) CM with uniform signaling (compare green circles with crosses and red curve).

\begin{figure}[!t] \centering 
\includegraphics[width=\columnwidth]{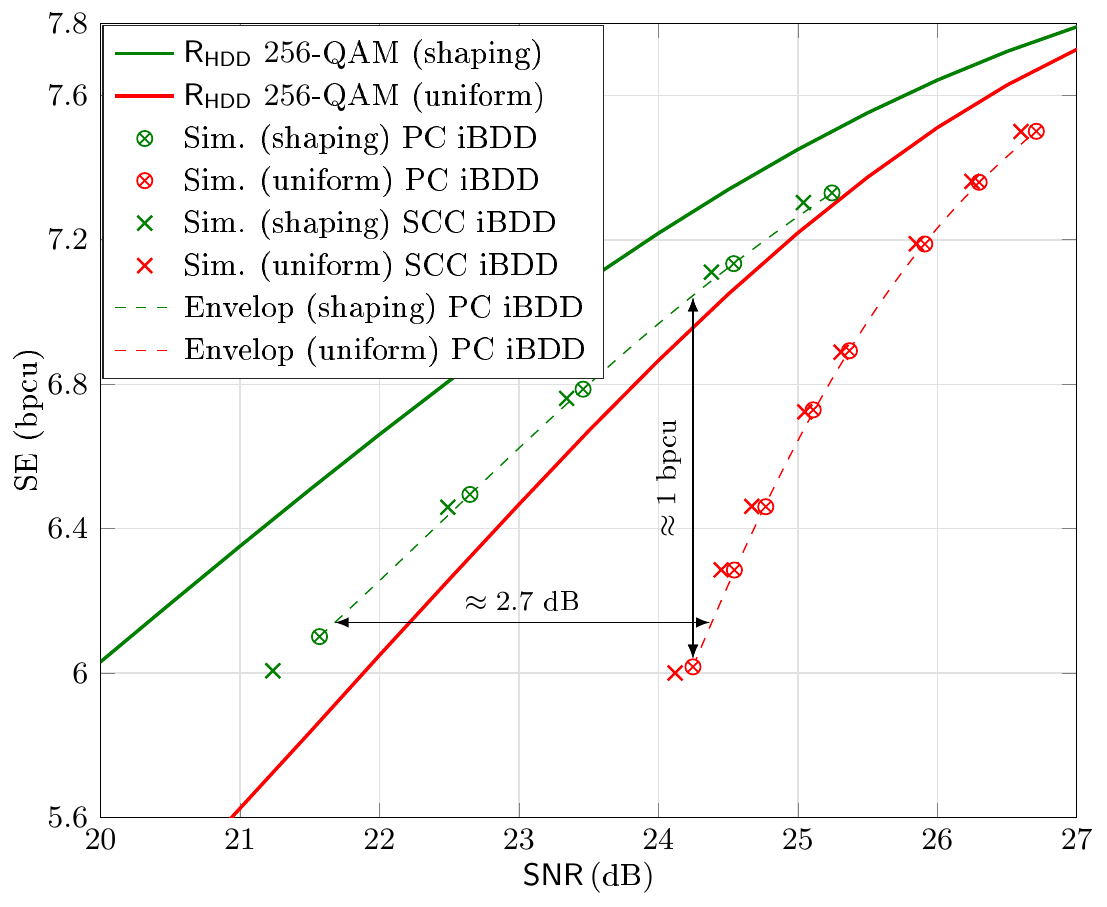}  
	\vspace{-4ex}
	\caption{Simulation results of PAS with PC and SCCs with parameters summarized in Table~\ref{table:feasibleparam} and \cite[Table.~II]{Sheikh2017Shap}, respectively, for $256$-QAM, and comparison with the corresponding achievable information rates.}  \vspace{-2ex}
	\label{16_ASK_sim2} 
\end{figure}

\begin{figure}[!t] \centering 
 \includegraphics[width=\columnwidth]{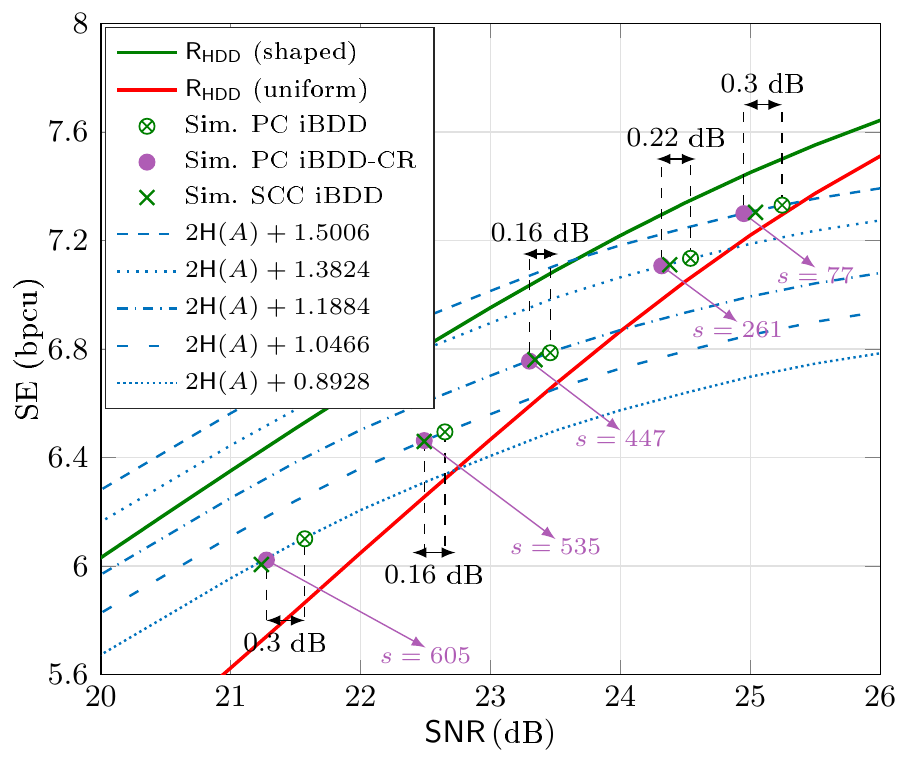}    
	\vspace{-4ex}
	\caption{Performance of the PAS with PC and iBDD-CR algorithm, PC with iBDD, and SCC with iBDD. Starting from top right, the shortening parameter for PC is $77$, $266$, $447$, $535$, and $605$, and shortening parameter for SCC is $63$, $247$, $431$, $519$, and $591$.}  \vspace{-2ex}
	\label{sim_total} 
\end{figure}
For the sake of comparison, we consider the same scenario for PAS and SCCs using a window decoder of size $7$ staircase blocks and $8$ decoding iterations, as considered in \cite[Sec.~VI]{Sheikh2017Shap}. In particular, to have a fair comparison between SCCs and PCs, we consider SCCs with $(v,t)=(10,3)$ and $s=63$, $274$, $431$, $519$, and $591$, which gives virtually the same code rates as those of the PCs considered in Fig.~\ref{16_ASK_sim2}. 
Comparing green crosses with green circles with crosses, one can see that for PCs the back-off SNR is larger than for SCCs, hence, PAS with PCs yields higher SEs than PAS with SCCs, at the cost of higher operating SNR. 

In Fig.~\ref{16_ASK_sim2}, we compare the performance of PAS with PCs and SCCs with uniform signaling. In particular, we consider PCs with $s=77$, $289$, $447$, $605$, $661$, $727$, $759$, and $797$. These PCs have roughly the same code rate as SCCs with $s=63$, $271$, $431$, $591$, $647$, $711$, $743$, and $783$ that are considered in \cite[Fig.~8]{Sheikh2017Shap}. As can be seen, the performance of both PAS with PCs and uniform signaling varies on an envelope (see green and red dashed curves in Fig.~\ref{16_ASK_sim2}), where the performance improvement of PCs with shaping over the uniform signaling reaches up to $2.7$ dB. Furthermore, comparing the SEs of PAS with PCs and uniform signaling, one can see that employing shaping provides up to $1$ bpcu SE improvement.
Comparing the simulation points with the corresponding achievable information rate (AIR) curve, it is clear that by reducing the code rate for both PCs and SCCs, the SNR gap between operating points and the AIR increases. It is worth to point out that this gap is always lower for the PAS with PCs compared to PCs with uniform signaling.

In Fig.~\ref{sim_total}, we compare the performance of PAS with PCs decoded using iBDD-CR \cite{optimal_decsheikh}, with that of PAS and both PCs and SCCs decoded using iBDD. Comparing the operating points of PAS with iBDD-CR and iBDD of PCs, an SNR gain of up to $0.3$ dB is observed. The improvement yielded by iBDD-CR is at the cost of computing and storing the channel LLRs compared to iBDD \cite{optimal_decsheikh}. 
Interestingly, PCs with iBDD-CR decoding and $s=77$, $261$, and $447$ provide  better operating points than that of SCCs with iBDD and same code rate ($s=63$, $247$, and $431$). Furthermore, the PC with iBDD-CR and $s=519$ provides the same performance as the corresponding SCC ($s=519$) with iBDD. Also, the PC with iBDD-CR and $s=605$ provides slightly worse performance compared to the corresponding SCC ($s=591$) with iBDD.

\section{Conclusion}\label{conclusion}

\SH{We considered probabilistic amplitude shaping with binary PCs and HDD based on iBDD of the component codes.} In particular, we addressed the parameter selection of the PC component codes and found the corresponding SE operating points. We showed that the performance of PAS with PCs is up to $2.7$ dB  better than that of the standard CM scheme with PCs and uniform signaling. Furthermore, we showed that PAS combined with the iBDD-CR decoding algorithm for PCs outperforms PAS with conventional iBDD for PCs, \SH{where the  performance improvement is up to $0.3$ dB. Interestingly, we observed that PAS with PCs and iBDD-CR can close the gap between PAS with iBDD of SCCs and PCs. Future work includes extending of iBDD-CR to SCCs, and  investigating the performance of PAS with iBDD-CR of SCCs.} 

We highlight that in this paper we used a pragmatic approach to evaluate the performance of PAS with PCs and iBDD-CR, by maximizing the AIR of the PAS with HDD. In principle, hybrid decoders (such as iBDD-CR) operate beyond the Hamming metric. Therefore, the operating point shown in this paper can be improved further by first deriving the AIR of the hybrid system and then optimizing the AIR. This analysis is left as future work.

\balance

\vspace{-0.5cm}


\begin{thebibliography}{10}
	\providecommand{\url}[1]{#1}
	\csname url@samestyle\endcsname
	\providecommand{\newblock}{\relax}
	\providecommand{\bibinfo}[2]{#2}
	\providecommand{\BIBentrySTDinterwordspacing}{\spaceskip=0pt\relax}
	\providecommand{\BIBentryALTinterwordstretchfactor}{4}
	\providecommand{\BIBentryALTinterwordspacing}{\spaceskip=\fontdimen2\font plus
		\BIBentryALTinterwordstretchfactor\fontdimen3\font minus
		\fontdimen4\font\relax}
	\providecommand{\BIBforeignlanguage}[2]{{%
			\expandafter\ifx\csname l@#1\endcsname\relax
			\typeout{** WARNING: IEEEtran.bst: No hyphenation pattern has been}%
			\typeout{** loaded for the language `#1'. Using the pattern for}%
			\typeout{** the default language instead.}%
			\else
			\language=\csname l@#1\endcsname
			\fi
			#2}}
	\providecommand{\BIBdecl}{\relax}
	\BIBdecl
	
	\bibitem{fujitsu2019}
	Fujitsu, ``{1FINITY T600 Transport Blade},'' Online:
	https://www.fujitsu.com/us/Images/1FINITY-T600-Data-Sheet.pdf.
	
	\bibitem{Elias1954}
	P.~Elias, ``Error-free coding,'' \emph{Trans. IRE Professional Group on Inf.\
		Theory}, vol.~4, no.~4, pp. 29--37, Sep. 1954.
	
	\bibitem{staircase_frank}
	B.~P. Smith, A.~Farhood, A.~Hunt, F.~R. Kschischang, and J.~Lodge, ``Staircase
	codes: {FEC} for 100 {G}b/s {OTN},'' \emph{IEEE/OSA J.\ Lightw.\ Technol.},
	vol.~30, no.~1, pp. 110--117, Jan. 2012.
	
	\bibitem{ITUT}
	``Forward error correction for high bit-rate {DWDM} submarine systems,'' ITU-T
	Recommendation G.975.1, 2004.
	
	\bibitem{400GZR}
	O.~I. Forum, ``Implementation agreement {400ZR},'' 2018.
	
	\bibitem{optimal_decsheikh}
	\BIBentryALTinterwordspacing
	A.~Sheikh, A.~{Graell i Amat}, G.~Liva, and A.~Alvarado, ``Refined reliability
	combining for binary message passing decoding of product codes,''
	\emph{arXiv}, 2020. [Online]. Available:
	\url{https://arxiv.org/abs/2006.00070}
	\BIBentrySTDinterwordspacing
	
	\bibitem{She19Tcom}
	A.~Sheikh, A.~{Graell i Amat}, and G.~Liva, ``Binary message passing decoding
	of product-like codes,'' \emph{IEEE Trans.\ Commun.}, vol.~67, no.~12, pp.
	8167--8178, Dec. 2019.
	
	\bibitem{She19}
	------, ``Binary message passing decoding of product codes based on generalized
	minimum distance decoding,'' in \emph{Proc. 53rd Annu. Conf. Inf. Sciences
		and Systems (CISS)}, Baltimore, MD, Mar. 2019.
	
	\bibitem{FougstedtiBDDSR2018}
	C.~Fougstedt, A.~Sheikh, A.~{Graell i Amat}, G.~Liva, and P.~Larsson-Edefors,
	``Energy-efficient soft-assisted product decoders,'' in \emph{Proc.\ Optical
		Fiber Commun. Conf. (OFC)}, San Diego, CA, USA, Mar. 2019.
	
	\bibitem{Montorsi2019}
	G.~{Montorsi}, ``Low complexity two-stage decoders for bawgn,'' in \emph{ICC
		2019 - 2019 IEEE International Conference on Communications (ICC)}, Shanghai,
	China, May 2019.
	
	\bibitem{Yibitflip}
	Y.~Lei, A.~Alvarado, B.~Chen, X.~Deng, Z.~Cao, J.~Li, and K.~Xu, ``Decoding
	staircase codes with marked bits,'' in \emph{Proc. Int. Symp. Turbo Codes and
		Iterative Inf. Processing (ISTC)}, Hong Kong, Dec. 2018.
	
	\bibitem{Ksch18}
	M.~Barakatain and F.~R. Kschischang, ``Low-complexity concatenated
	{LDPC}-staircase codes,'' \emph{IEEE/OSA J.\ Lightw.\ Technol.}, vol.~36,
	no.~12, pp. 2443--2449, Jun. 2018.
	
	\bibitem{Liga2019bit}
	G.~Liga, A.~Sheikh, and A.~Alvarado, ``A novel soft-aided bit-marking decoder
	for product codes,'' in \emph{Proc.\ European Conf. Optical Communications
		(ECOC)}, Dublin, Ireland, Sep. 2019.
	
	\bibitem{Calderbank_1990}
	A.~R. Calderbank and L.~H. Ozarow, ``Nonequiprobable signaling on the
	{Gaussian} channel,'' \emph{IEEE Trans.\ Inf.\ Theory}, vol.~36, no.~4, pp.
	726--740, Jul. 1990.
	
	\bibitem{Forney_1992}
	{G. D. Forney, Jr.}, ``Trellis shaping,'' \emph{IEEE Trans.\ Inf.\ Theory},
	vol.~38, no.~2, pp. 281--300, Mar. 1992.
	
	\bibitem{georg_tcom}
	G.~B\"ocherer, F.~Steiner, and P.~Schulte, ``Bandwidth efficient and
	rate-matched low-density parity-check coded modulation,'' \emph{IEEE Trans.\
		Commun.}, vol.~63, no.~12, pp. 4651--4665, Dec. 2015.
	
	\bibitem{Buchali_2016}
	{F. Buchali}, {F. Steiner}, {G. B\"ocherer}, {L. Schmalen}, {P. Schulte}, and
	{W. Idler}, ``Rate adaptation and reach increase by probabilistically shaped
	64-{QAM}: An experimental demonstration,'' \emph{IEEE/OSA J.\ Lightw.\
		Technol.}, vol.~34, no.~7, pp. 1599--1609, Apr. 2016.
	
	\bibitem{Sheikh2017Shap}
	A.~Sheikh, A.~{Graell i Amat}, G.~Liva, and F.~Steiner, ``Probabilistic
	amplitude shaping with hard decision decoding and staircase codes,''
	\emph{IEEE/OSA J.\ Lightw.\ Technol.}, vol.~36, no.~9, pp. 1689--1697, May
	2018.
	
	\bibitem{boherer20191}
	G.~{B\"ocherer}, P.~{Schulte}, and F.~{Steiner}, ``Probabilistic shaping and
	forward error correction for fiber-optic communication systems,''
	\emph{IEEE/OSA J.\ Lightw.\ Technol.}, vol.~37, no.~2, pp. 230--244, Jan.
	2019.
	
	\bibitem{Cho2019}
	J.~{Cho} and P.~J. {Winzer}, ``Probabilistic constellation shaping for optical
	fiber communications,'' \emph{IEEE/OSA J.\ Lightw.\ Technol.}, vol.~37,
	no.~6, pp. 1590--1607, Mar. 2019.
	
	\bibitem{Poggiolini2015}
	P.~{Poggiolini}, G.~{Bosco}, A.~{Carena}, V.~{Curri}, Y.~{Jiang}, and
	F.~{Forghieri}, ``A simple and effective closed-form {GN} model correction
	formula accounting for signal non-gaussian distribution,'' \emph{IEEE/OSA J.\
		Lightw.\ Technol.}, vol.~33, no.~2, pp. 459--473, Jan. 2015.
	
	\bibitem{bocherer2017}
	\BIBentryALTinterwordspacing
	G.~B\"{o}cherer, ``Achievable rates for probabilistic shaping,'' V5, May, 2017.
	[Online]. Available: \url{http://arxiv.org/abs/arXiv:1707.01134.}
	\BIBentrySTDinterwordspacing
	
	\bibitem{shapingGeorg}
	P.~Schulte and G.~B\"ocherer, ``Constant composition distribution matching,''
	\emph{IEEE Trans.\ Inf.\ Theory}, vol.~62, no.~1, pp. 430--434, Jan. 2016.
	
	\bibitem{Fehenberger2018multi}
	T.~{Fehenberger}, D.~S. {Millar}, T.~{Koike-Akino}, K.~{Kojima}, and
	K.~{Parsons}, ``Multiset-partition distribution matching,'' \emph{IEEE
		Trans.\ Commun.}, vol.~67, no.~3, pp. 1885--1893, Mar. 2019.
	
	\bibitem{Justesen2011}
	J.~Justesen, ``Performance of product codes and related structures with
	iterated decoding,'' \emph{IEEE Trans.\ Commun.}, vol.~59, no.~2, pp.
	407--415, Feb. 2011.
	
	\bibitem{h_braided}
	Y.~Y. Jian, H.~D. Pfister, K.~R. Narayanan, R.~Rao, and R.~Mazahreh,
	``Iterative hard-decision decoding of braided {BCH} codes for high-speed
	optical communication,'' in \emph{Proc.\ IEEE Global Telecommun.\ Conf.
		(GLOBECOM)}, Atlanta, GA, Dec. 2013.
	
\end{thebibliography}


\end{document}